\documentclass[12pt]{article}
\usepackage{color}
\usepackage{amsmath,amsfonts,graphicx,yfonts,mathabx,braket,mathtools,subfigure,slashed}
\usepackage[nosort]{cite}
\newcommand{\be}{\begin{equation}}
\newcommand{\ee}{\end{equation}}
\newcommand{\bea}{\begin{eqnarray}}
\newcommand{\eea}{\end{eqnarray}}
\renewcommand{\title}[1]{\vbox{\center\LARGE{#1}}\vspace{3mm}}
\renewcommand{\author}[1]{\vbox{\center#1}\vspace{3mm}}
\newcommand{\address}[1]{\vbox{\center\em#1}}
\newcommand{\email}[1]{\vbox{\center\tt#1}\vspace{3mm}}

\begin{document}
\begin{titlepage}
\begin{center}
\rightline{\tt}
\vskip 2.5cm
\title{{\bf Supersymmetric integrable theories \\ from no particle production}}
\vskip .6cm
\author{Carlos Bercini and Diego Trancanelli}
\vskip -.5cm 
\address{Institute of Physics, University of S\~ao Paulo\\ 05314-970 S\~ao Paulo, Brazil}
\vskip -.1cm 
\email{cbercini, dtrancan@if.usp.br}
\end{center}
\vskip 3cm

\abstract{\noindent  
We consider a theory of scalar superfields in two dimensions with arbitrary superpotential. By imposing no particle production in tree-level scattering, we constrain the form of the admissible interactions, recovering a supersymmetric extension of the sinh-Gordon model.  
}
\vfill
\end{titlepage}


Recent years have witnessed a revival \cite{Paulos:2016fap} of the old S-matrix bootstrap program~\cite{old,old1}, which is the attempt to solve massive quantum field theories solely from imposing basic consistency principles, such as analyticity, unitarity, and crossing symmetry of their scattering matrices. One very simple arena where this can be carried out is the case of two-dimensional integrable theories, whose S-matrix is severely constrained and characterized by factorization and absence of particle production; see e.g. the nice reviews in \cite{Dorey:1996gd,Mussardo:2010mgq,Bombardelli:2016scq}.

One can imagine of ``rediscovering'' the existence of such integrable theories by imposing the absence of particle production in scattering processes and thus checking which kind of interactions are admissible \cite{pedro}. The scope of this short note is to carry out this procedure for a simple supersymmetric model in two dimensions, the scalar ${\cal N}=(1,1)$ superfield. Writing a generic superpotential for this field and imposing that no particles be produced introduces a recursion relation among the couplings of generic vertices and allows to recover the ${\cal N}=1$ supersymmetric sinh-Gordon model, which is known to be integrable from the explicit construction of  infinite conserved charges \cite{DiVecchia:1977nxl,Ferrara:1978jv,Shankar:1977cm,Sengupta:1985tk}.\footnote{Absence of particle production in the bosonic sine-Gordon model was proven in \cite{Arefeva:1974bk}.}

Our starting point is an ${\cal N}=(1,1)$ scalar superfield $\Phi$ in two dimensions with generic superpotential $W(\Phi)$. This superfield consists of a real scalar field $\phi(x)$, a two-component Majorana spinor $\psi(x)$, and an auxiliary field $F(x)$; see e.g. \cite{Hori:2003ic}. The Lagrangian for these components is
\begin{equation}
    \mathcal{L} = \frac{1}{2}\partial_\mu\phi\partial^\mu\phi + \bar{\psi}i\slashed{\partial}\psi - \frac{1}{2}(W'(\phi))^2 -W''(\phi)\bar{\psi}\psi\,,
    \label{onshellLagrangian}
\end{equation}
where the auxiliary field $F(x)$ has been eliminated by imposing its equation of motion, $F=-W'(\phi)$. We take $\gamma^\mu = \{ \sigma_2,-i\sigma_1 \}$.
In two dimensions, a scalar field has scaling dimension zero, so that arbitrary powers of $\phi$ can be included in $W(\phi)$ without spoiling renormalizability:
\begin{equation}
    W(\phi) =  \frac{1}{2}M\phi^2 + \sum_{n=3}^{\infty}\frac{\lambda_n}{n!}\phi^n\,.
    \label{superpotential}
\end{equation}
This results in the Lagrangian
\begin{equation}
    \mathcal{L} = \frac{1}{2}\partial_\mu\phi\partial^\mu\phi - \frac{1}{2}M^2\phi^2 + \bar{\psi}i\slashed{\partial}\psi - M\bar{\psi}\psi - \sum_{n=3}^{\infty}\frac{\Lambda_n}{n!}\phi^n - \sum_{n=3}^{\infty}\frac{\lambda_n}{(n-2)!}\phi^{n-2}\bar{\psi}\psi\,,
    \label{initlag}
 \end{equation}
with scalar couplings
\begin{equation}
    \Lambda_n = Mn\lambda_n + \frac{n!}{2}\sum_{i=3}^n\sum_{j=3}^n\frac{\lambda_i}{(i-1)!}\frac{\lambda_j}{(j-1)!}\delta_{i+j-2,n}.
    \label{biglambda}
\end{equation}

We now consider tree-level scattering and impose that no particles be produced in these processes. This will introduce recursion relations in the couplings, which can be solved to find the generic expression for $\lambda_n$ and the potential in (\ref{superpotential}). It is sufficient to look at scattering amplitudes with just two initial states: ${\cal M}_{2\to n}$. Our convention is that all particles are taken to be incoming, with the understanding that in the end all but two particles are crossed to outgoing.

It is convenient to go to light-cone coordinates and write the momentum of the $i$-th particle $(p_i^+,p^-_i)=(p^0_i+p^1_i,p^0_i-p^1_i)$ in terms of a single real parameter $a_i$ as
\be
p_i=(Ma_i,M/a_i)\,,
\ee
which guarantees $p^+_i p^-_i=M^2$. From now on we set $M=1$ without loss of generality. Conservation of energy and momentum is expressed as 
\be
\sum_i a_i=\sum_i \frac{1}{a_i}=0\,.
\label{momcons}
\ee
The scalar and fermion propagators carrying momentum $p=\sum_i p_i$ are given, respectively, by
\be
G(p)=\frac{1}{p^2-1}=\frac{1}{\left(\sum_i a_i\right)\left(\sum_j 1/a_j\right)-1}\,,
\ee 
and
\be
\label{fermprop}
 \slashed{G}(p)\equiv \frac{\slashed{p}+1}{p^2-1}=G(p)\left(\begin{array}{cc} 1 & -i \sum_i 1/a_i \\ i \sum_i a_i & 1\end{array}\right)\,.
\ee

\subsubsection*{The bosonic sector}

\begin{figure}[h!]
\begin{center}
    \includegraphics[scale=0.4]{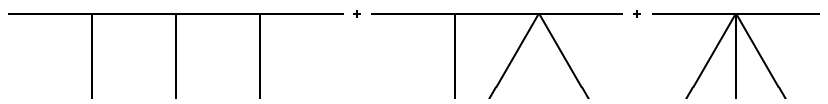}
    \end{center}
    \caption{\small Tree-level diagrams contributing to ${\cal M}_{2\textrm{b} \to 3\textrm{b}}$.}
    \label{FigureM23}
\end{figure}

Let us focus for the moment on the purely bosonic case, which is also studied in \cite{pedro}. The simplest scattering process with particle production is ${\cal M}_{2\textrm{b} \to 3\textrm{b}}$, which is evaluated at tree level by computing the diagrams in Fig.~\ref{FigureM23}.

The key observation \cite{Dorey:1996gd} is that if one could set the sum of the first two diagrams in the figure to be constant, for arbitrary configurations of momenta obeying (\ref{momcons}), these first two diagrams could be eliminated by the third one, by simply tuning $\Lambda_5$ to minus that constant.  One can check that this strategy can indeed be implemented! Imposing $\Lambda_4=3 \Lambda_3^2$ guarantees that the sum of the first two diagrams be a constant, which happens to be equal to $-5\Lambda^3_3$. This process of particle creation can then be eliminated by tuning $\Lambda_5=5\Lambda^3_3$. 

Of course, the next step would be to generalize this procedure for amplitudes with an arbitrary number of final bosons, in order to find expressions for all the couplings in terms of $\Lambda_3$. This can be done recursively, using the factorization of the amplitudes and a clever choice of momenta. Note in fact that all tree-level diagrams for $n$ particles, except for the constant one equal to $\Lambda_n$, can be factorized in a left blob and a right blob connected by a propagator
\be
G_{L\to R}=\frac{1}{\left(\sum_{i\in L}a_i\right)\left(\sum_{j\in L}1/a_j\right)-1}\,.
\ee
One can also show from the analyticity properties of the amplitudes \cite{pedro} that the sum of all diagrams remains constant, even for a generic number of exernal particles. This allows to pick a convenient choice of momenta which simplifies the evaluation of the recursive relations among the $\Lambda_n$'s. This convenient choice turns out to be \cite{pedro}
\begin{equation}
     a = \{a_1(x),1,x,x^2,\dots,x^{n-3},a_n(x)\}\,,
     \label{momentapick}
\end{equation}
with $a_i(x)$ and $a_n(x)$ being determined by (\ref{momcons}). In the limit of $x\to\infty$, 
\begin{equation}
     a_1(x) = -1 + \mathcal{O}(1/x)\,,\qquad 
     a_n(x) = -x^{n-3}\left(1 + \mathcal{O}(1/x)\right)\,,
\end{equation}
and one can check that
\begin{equation}
 G_{L\to R} = \left\{
     \begin{array}{rl}
          -1, & \qquad \textnormal{if } a_j^L = (a_1,a_2,\dots,a_k), a_j^R = (a_{k+1},a_{k+2},\dots,a_n)  \\
          0, & \qquad \textnormal{otherwise} \\
     \end{array}
     \right.\,,
     \label{EqScalarLR}
\end{equation}
that is, only the diagrams where the particles are ordered contribute. To see this, it is sufficient to evaluate a few cases. For example, the set $a_j^L = \{a_2,a_3\}$ gives a vanishing propagator and it is easy to see that any other set of two or more momenta not including $a_1$ goes to zero as well. Similarly, the set $a_j^L = \{a_1,a_{i\ge 3}\}$ also gives zero. On the other hand, the set $a_j^L = \{a_1,a_2\}$ yields $-1$. By adding $a_{i\ge 4}$ to this set, we get zero again. The only case left to analyze is if an ordered set of any size continues to converge to $-1$, which it does.

\begin{figure}[h!]
\begin{center}
\begin{tabular}{cc}
\setlength{\unitlength}{1cm}
\hspace{-0.9cm}
    \includegraphics[scale=0.4]{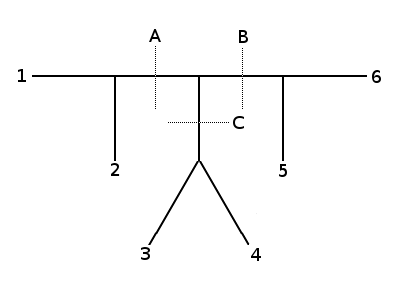}
    \end{tabular}
\end{center}
    \caption{\small 
    This diagram seems to be non-vanishing since it is ordered for the cuts $A$ and $B$. However, for the cut $C$ it is $a_j^L = \{a_3,a_4 \}$, which implies that this diagram does in fact vanish. In general, only (ordered) line-type diagrams survive the large $x$ limit. 
    }
    \label{FigDiagLR1}
\end{figure}

An important consequence of (\ref{EqScalarLR}) is that only ordered line-type diagrams survive, as shown pictorially in Fig.~\ref{FigDiagLR1} for the case of six particles.

\begin{figure}[h!]
\begin{center}
    \includegraphics[scale=0.4]{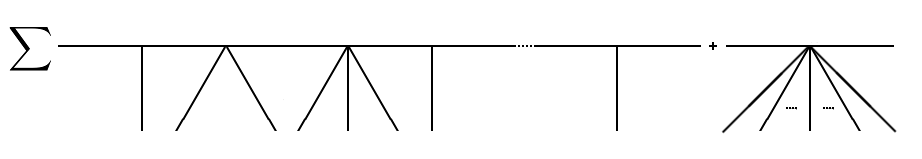}
    \end{center}
    \caption{\small The ${\cal M}_{2\textrm{b} \rightarrow (n-2)\textrm{b}}$ scattering. The sum is over all ordered line-typed diagrams.}
    \label{FigMNLineType}
\end{figure}

After these considerations, one is ready to compute the amplitude $\mathcal{M}_{2\textrm{b} \rightarrow (n-2)\textrm{b}}$, schematically shown in Fig.~\ref{FigMNLineType}.
It is useful to start thinking about the diagram that has $\Lambda_{n-k}$ as its rightmost vertex. This diagram factorizes into the form $\mathcal{M}_{2\textrm{b} \rightarrow k\textrm{b}}\cdot \Lambda_{n-k}$. Imposing $\mathcal{M}_{2\textrm{b} \rightarrow k\textrm{b}}$ to vanish, one gets the recursion relation
\begin{equation}
    \mathcal{M}_{2\textrm{b} \rightarrow (n-2)\textrm{b}} = -\Lambda_3\Lambda_{n-1} + (\Lambda_3)^2\Lambda_{n-2} - \Lambda_4\Lambda_{n-2} + \Lambda_n = 0\,.
    \label{RecursionRel}
\end{equation}
This can be solved by setting $\Lambda_n = \gamma^n$, writing a generic combination of the two roots, and imposing consistency with $\Lambda_4=3\Lambda^2_3$. One finds
\begin{equation}
    \Lambda_n = \frac{\Lambda_3^{n-2}}{6}(2(-1)^n + 2^n)\,.
    \label{LambdanBD}
\end{equation}
Plugging this in the potential in (\ref{initlag}) and resumming, one ends up with the Bullough-Dodd potential \cite{Dodd:1977bi,Braden:1989bu,Braden:1991vz}, a known integrable model in two dimensions:
\begin{equation}
    \mathcal{L} = \frac{1}{2}\partial_\mu \phi \partial^\mu \phi - \frac{1}{6\Lambda_3^2}\left( 2 e^{-\Lambda_3\phi} + e^{2\Lambda_3\phi} -3 \right)\,.
\end{equation}

\begin{figure}[h!]
\begin{center}
    \includegraphics[scale=0.4]{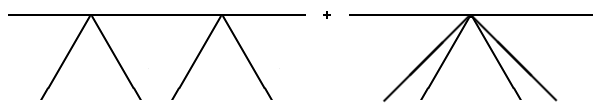}
    \end{center}
    \caption{\small Tree-level diagrams contributing to ${\cal M}_{2\textrm{b} \to 4\textrm{b}}$ in a $\mathbb{Z}_2$-invariant theory.}
    \label{FigureM24}
\end{figure}

One can repeat this exercise, imposing extra symmetries, for example invariance under $\phi \to -\phi$. Odd powers of the expansion must vanish, so that $\Lambda_\textrm{odd}=0$. The amplitude $\mathcal{M}_{2\textrm{b} \rightarrow 3\textrm{b}}$ is now trivially zero, the first non-trivial process being $\mathcal{M}_{2\textrm{b} \rightarrow 4\textrm{b}}$, shown in Fig.~\ref{FigureM24}.
This amplitude vanishes provided $\Lambda_6 = \Lambda_4^2$. Going through the same steps as above for generic diagrams, one finds the recursion
\begin{equation}
    -\Lambda_{n-2}\Lambda_4 + \Lambda_n = 0\,,
    \label{RecusiRelCoupEven}
\end{equation}
which is solved by 
\be
\Lambda_n=\Lambda_4^{\frac{n-2}{2}}\,.
\label{SUSYEvenconstraintBL}
\ee 
Resumming the potential results in the sinh-Gordon potential
\begin{equation}
     \mathcal{L} = \frac{1}{2}\partial_\mu \phi \partial^\mu \phi - \frac{1}{\Lambda_4}(\cosh{(\sqrt{\Lambda_4}\phi)}-1)\,,
\end{equation}
another known integrable model in two dimensions.

\subsubsection*{Including the fermions}

We now move on to the supersymmetric case. We can extend the considerations above to the fermionic propagator in (\ref{fermprop}). In the limit $x \rightarrow \infty $ for the momenta in (\ref{momentapick}), this is such that
\begin{equation}
 \slashed{G}_{L\to R}(p_1,\ldots) \rightarrow 
\left( \begin{array}{cc}
-1 & 0 \\ -i x^{\textrm{max}\{i\in L\}-2} & -1 
\end{array}
\right)\,,
     \label{FermionicProp0}
\end{equation}
when the particles are ordered and
\begin{equation}
 \slashed{G}_{L\to R}(p_1,\ldots) \rightarrow 
\slashed{q},
     \label{FermionicProp}
\end{equation}
with $q=(-1,0)$, otherwise. Ordered diagrams involving $\slashed{G}_{L\to R}$ do not appear at tree level, so that (\ref{FermionicProp0}) will not be necessary.

Initially, we restrict our attention to theories preserving a $\mathbb{Z}_2$-symmetry. These are invariant under $\phi\to -\phi$ and have $\lambda_\textrm{odd}=0$, which implies $\Lambda_\textrm{odd}=0$. The processes which involve bosons only are the same as the real scalar theories studied above. One finds the recursion relation (\ref{SUSYEvenconstraintBL}) for the scalar couplings, which, using equation (\ref{biglambda}), translates into an identical relation for the $\lambda_n$'s appearing in the superpotential
\begin{equation}
    \lambda_n = \lambda_4^{\frac{n-2}{2}}\,.
    \label{SUSYEvenconstraintSL}
\end{equation}
Performing the sum, one finds that
\begin{equation}
    W(\phi) = \frac{1}{\lambda_4}\left(\cosh{\sqrt{\lambda_4}\phi}-1\right)\,,
\end{equation}
which is the superpotential of the ${\cal N}=1$ supersymmetric sinh-Gordon model \cite{DiVecchia:1977nxl,Ferrara:1978jv,Shankar:1977cm,Sengupta:1985tk}. 

In order to complete the derivation, one must prove that this particular choice of couplings imply no particle production in the processes which include fermions.

\paragraph{2 fermions into 4 bosons.}
First, we consider the simplest possible scattering process where particle production could occur, namely $\bar{\psi}\psi \rightarrow \phi^4$, and show that the corresponding amplitude ${\cal M}_{\textrm{2f} \to \textrm{4b}}$ is zero, independently of the choice of momenta. 
\begin{figure}
\begin{center}
\begin{tabular}{c}
\setlength{\unitlength}{1cm}
\hspace{-0.9cm}
    \includegraphics[scale=0.5]{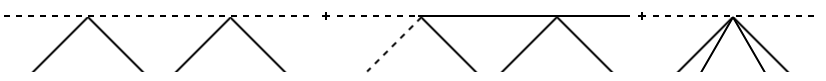} \\
  \hskip 0.9cm  (a) \hskip 5.2cm (b) \hskip 3.5cm (c)
    \end{tabular}
\end{center}
    \caption{\small Diagrams contributing to the process $\bar{\psi}\psi \rightarrow \phi\phi\phi\phi$. Solid lines represent bosons and dashed lines represent fermions. 
    }
    \label{FigureSUSYM2f4b}
\end{figure}
The three diagrams in Fig.~\ref{FigureSUSYM2f4b} contribute to this process. Their evaluation is simplified enormously by considering the choice of momenta in (\ref{momentapick}) above.

The process has only two fermions, with associated momenta $p_1$ and $p_2$. Since the diagrams are evaluated at tree level, the fermionic propagator will never be ordered. Whenever a fermionic propagator appears in a diagram, it can be replaced by $\slashed{q}$. The amplitude ${\cal M}_{\textrm{2f} \to \textrm{4b}}$ will never display any divergence in this limit: the fermionic propagator cannot diverge and the bosonic propagator goes to zero or $-1$. If any term of a given diagram goes to zero, it can be discarded from the very beginning. 

In the limit $x \rightarrow \infty$, we evaluate the three classes of diagrams in Fig.~\ref{FigureSUSYM2f4b} as
\bea
&&        \textrm{(a)} \quad \rightarrow  \quad\bar{v}_2 (6\lambda_4^2 \slashed{q} )u_1\,,\cr
 &&       \textrm{(b)} \quad \rightarrow  \quad\bar{v}_2(-\lambda_4\Lambda_4)u_1 = \bar{v}_2(-4\lambda_4^2)u_1\,, \cr
 &&       \textrm{(c)}  \quad\rightarrow \quad \bar{v_2}(\lambda_6)u_1\,.
\eea
Here we employ the usual conventions for the external fermionic particles: $u$ ($\bar u$) denotes the initial (final) fermions, whereas $v$ ($\bar v$) is for the final (initial) antifermions.   

Diagram (c) is trivially evaluated, diagram (b) has only one non-vanishing configuration, whereas diagram (a) has six contributing configurations, resulting in the amplitude
\begin{equation}
    {\cal M}_{\textrm{2f} \to \textrm{4b}} \rightarrow \bar{v}_2 (\lambda_4^26\slashed{q} + \lambda_6 - 4\lambda_4^2)u_1 = \lambda_4^2\bar{v}_2 (6\slashed{q} + \beta)u_1 \,,
\end{equation}
with $ \beta = \frac{1}{\lambda_4^2}(\lambda_6 -4\lambda_4^2)$. In order for the average modulus square $\overline{|  {\cal M}_{\textrm{2f} \to \textrm{4b}}|^2}$ to vanish, it must be $\beta = -3$, which gives the correct constraint for the couplings: $\lambda_6 = \lambda_4^2$, consistently with (\ref{SUSYEvenconstraintSL}).

\begin{figure}[h!]
\begin{center}
    \includegraphics[scale=0.45]{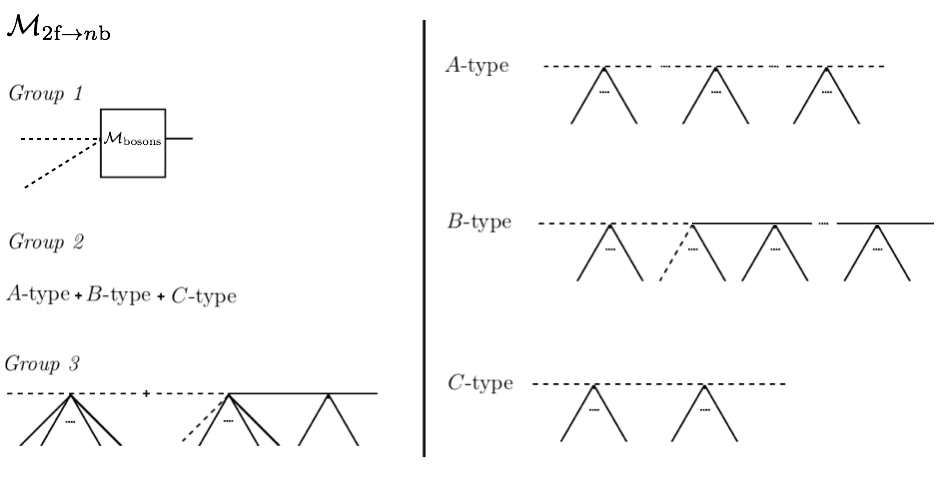}
    \end{center}
    \caption{\small Diagrams contributing to the process $\psi\bar{\psi} \rightarrow \phi^n$.}
    \label{FigureSUSYM2fnbGeral}
\end{figure}

\paragraph{2 fermions into $n$ bosons.}
The result above can be generalized to a general process involving two fermions going into $n$ bosons: $\psi\bar{\psi} \rightarrow \phi^n$, where $n$ is even by symmetry. All the diagrams are divided into three groups, schematically shown in Fig.~\ref{FigureSUSYM2fnbGeral}.

Diagrams in group 1 are essentially bosonic diagrams with an extra fermionic vertex attached to the first boson. They will always be vanishing, by the constraint imposed on bosonic diagrams above. Group 3 consists of two diagrams, which can be trivially evaluated to
\begin{equation}
    \bar{v}_2(\lambda_{n+2})u_1+ \bar{v}_2(\lambda_n(-1)\Lambda_4)u_1= \bar{v}_2(\lambda_{n+2} - 4\lambda_4^{n/2})u_1\,.
\end{equation}

Finally, group 2 is divided into three types of diagrams, which we dub of the $A$-type, $B$-type, and $C$-type. $A$-type diagrams have two or more fermionic propagators that will not be ordered, implying that the only surviving contribution will be proportional to $\slashed{q}\slashed{q}\dots\slashed{q}$. In the evaluation of $\overline{|{\cal M}_{\textrm{2f} \to n\textrm{b}}|^2}$ however, anything with at least two $\slashed{q}$'s will be proportional to $q \cdot q = 0$. Therefore, these diagrams do not contribute.

\begin{figure}
\begin{center}
    \includegraphics[scale=0.4]{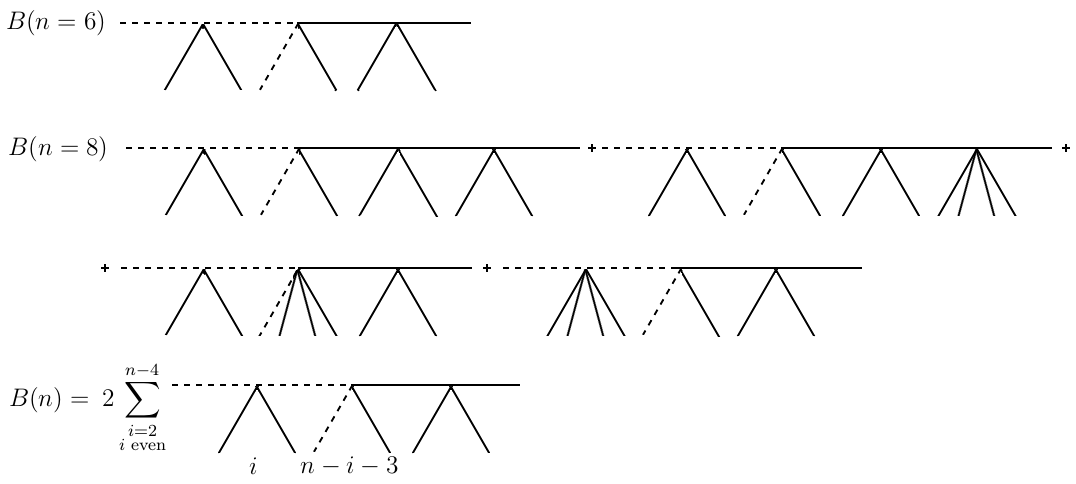}
    \end{center}
    \caption{\small $B$-type diagrams for $n=6,8$ and also a generalization for any $n$. The numbers between the external legs represent the number of bosons in the vertex.}
    \label{FigureSUSYM2fnbBtype}
\end{figure}
$B$-type diagrams are detailed in Fig.~\ref{FigureSUSYM2fnbBtype} and start appearing when $n=6$. They have only one fermionic propagator. Looking for example at the case $n=8$, it is easy to see that a cancellation takes place between the last terms due to the constraint imposed on the bosonic scattering, $\Lambda_6 = \Lambda_4^2$. This can be checked to happen also at arbitrary $n$, resulting in an effective expression given in the last line of Fig.~\ref{FigureSUSYM2fnbBtype}. The factor of 2 comes from an exchange of the fermions that can always be made in this kind of diagrams. Anything with a different arrangement of the right-most legs will contribute to the bosonic scattering (of $n-2$ bosons), already shown to be zero. In the end, the contribution of these diagrams to the amplitude is
\begin{equation}
    B(n) \rightarrow 2 \sum_{\substack{i = 2 \\ i \textnormal{ even}}}^{n-4}\lambda_{i+2}\lambda_{n-i}(-1)\Lambda_4\bar{v}_2\, b(n,i)\slashed{q}\, u_1 \,.
\end{equation}
The minus sign comes from the ordered bosonic propagator and $b(n,i)$ is the number of permutations which leave the last three legs with momenta ($n+2,n+1,n$) and give an ordered bosonic propagator. For each particular configuration, a propagator will depend on the $i$-th momentum, but since the last three must be fixed to get a non-zero contribution, $b(n,i)$ is the ways we can arrange $n-3$ objects into $i$ slots, without worrying about their order, so that
\begin{equation}
    B(n) \rightarrow  -8\lambda_4^{n/2} \sum_{\substack{i = 2 \\ i \textnormal{ even}}}^{n-4} \bar{v}_2\frac{(n-3)!}{i!(n-3-i)!} \slashed{q} u_1\,.
\end{equation}

\begin{figure}
\begin{center}
    \includegraphics[scale=0.4]{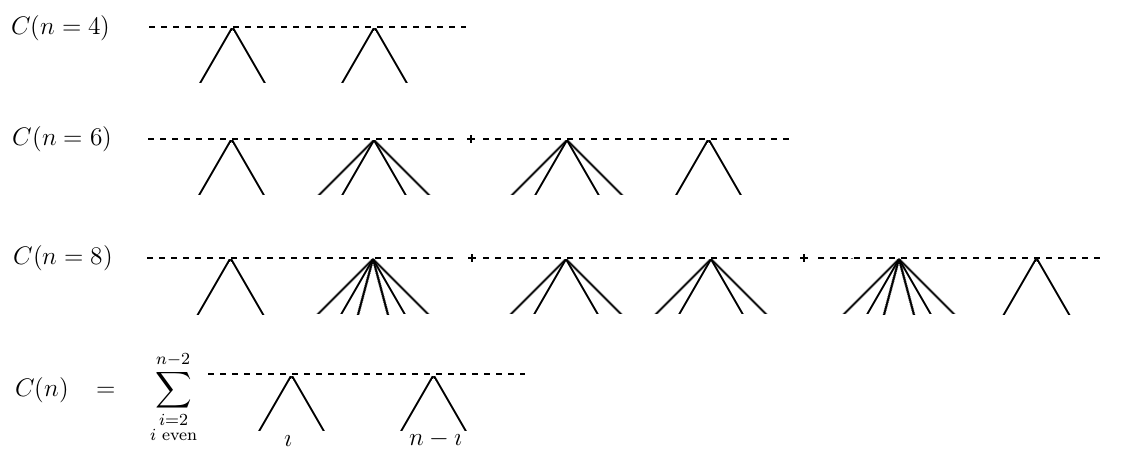}
    \end{center}
    \caption{\small $C$-type diagrams for $n=4,6,8$ and also a generalization for any $n$. The numbers between the external legs represent the number of bosons in the vertex.}
    \label{FigureSUSYM2fnbCtype}
\end{figure}
The general form of a $C$-type diagram is shown in Fig.~\ref{FigureSUSYM2fnbCtype}, remembering that only line-type diagrams will have an ordered bosonic propagator, potentially contributing to the amplitude. Each fermionic propagator can be split into a diagonal and off-diagonal part. Since the momenta are always out of order, the diagonal part will go to zero, and the off-diagonal one will go to $\slashed{q}$. This will happen for every permutation in the diagram, each one contributing in the same manner. The expression for $C(n)$ can be written as follows
\begin{equation}
    C(n) \rightarrow \sum_{\substack{i = 2 \\ i \textnormal{ even}}}^{n-2}\lambda_{i+2}\lambda_{n-i+2}\bar{v}_2\, c(n,i)\slashed{q}\, u_1\,.
\end{equation}
The number of permutations $c(n,i)$ is associated with the diagram of $i$ bosonic legs in the first vertex and $n-i$ bosonic legs in the second vertex. The amplitude associated with one configuration has a propagator that depends on the $i$-th momentum, so that $c(n,i)$ is the number of ways we can arrange $n$ objects into $i$ slots, irrespectively of their order:
\begin{equation}
    C(n) \rightarrow  \lambda_4^{n/2}\sum_{\substack{i = 2 \\ i \textnormal{ even}}}^{n-2}\bar{v}_2\frac{n!}{i!(n-i)!)}\slashed{q} u_1
\end{equation}
To write this expression we used a recursion, by assuming that the relation is satisfied to $\lambda_n$ in order to check if the result is correct for $\lambda_{n+2}$.

Summing $B$-type and $C$-type diagrams, we end up with the full contribution to the group 2 diagrams
\begin{equation}
     B(n)+C(n) = \lambda_4^{n/2}\bar{v}_2\alpha\slashed{q}u_1\,,
\end{equation}
where
\begin{equation}
    \alpha = \sum_{\substack{i = 2 \\ i \textnormal{ even}}}^{n-2}\frac{n!}{i!(n-i)!)} -8\sum_{\substack{i = 2 \\ i \textnormal{ even}}}^{n-4} \frac{(n-3)!}{i!(n-3-i)!} = 6\,.
\end{equation}

The amplitude of the process of two fermions going to $n$ bosons is given by the sum of these groups:
\begin{equation}
    {\cal M}_{\textrm{2f} \to n\textrm{b}} = \bar{v}_2 \left( \lambda_4^{n/2}6\slashed{q} +\lambda_{n+2} - 4\lambda_4^{n/2} \right) u_1 
    = \lambda_4^{n/2}\bar{v}_2(6\slashed{q}+\beta)u_1\,,
\end{equation}
where
\begin{equation}
    \beta = \frac{1}{\lambda_4^{n/2}}(\lambda_{n+2}-4\lambda_4^{n/2})\,.
\end{equation}
This amplitude is zero only if $\beta = -3$, which implies the same recursive relation as in (\ref{SUSYEvenconstraintSL}). Note that for the fermions the cancellation is at the level of the squared amplitude, not just at the level of the sum of the diagrams, as in the bosonic case.

\paragraph{2 bosons into $n$ fermions} The next class of diagrams left to be checked is two bosons going into four or more fermions. As explicitly show in \cite{Sengupta:1985tk}, the process of two bosons going into four fermions is given by one diagram that vanishes trivially. 
To extend this to $n$ fermions, we need to pick the set of momenta (\ref{momentapick}) and note that in the large $x$ limit we have
\be
\overline{|{\cal M}_{\textrm{2b} \to n\textrm{f}}|^2} \sim x^{(n-2)!-((n-2)!!)^2} \to 0\,.
\ee 
This happens just by the conservations laws (\ref{momcons}) and does not impose new constraints on the coupling constants.

\paragraph{2 particles into $n$ particles} Finally, with the ingredients derived so far, it is possible to show that all other processes in the theory involving two initial particles and more than two final particles are zero. To show this, one can think recursively and use the following facts:
\begin{itemize}
\item[{\it (i)}] One can flip signs in the set of momenta in (\ref{momentapick}) and exchange what we call ingoing and outgoing particles.
\item[{\it (ii)}] In the limit of $x$ going to infinity, only line-type diagrams will contribute and one can factorize the new cases into the cases which have already been shown to vanish.
\end{itemize}
Since we showed that $\psi\bar{\psi} \to \phi^4$ has vanishing amplitude, by {\it (i)} we are automatically guaranteed that $\phi\phi \to \phi^2\psi\bar{\psi}$ also vanishes, just like $\phi\psi \to \phi^3\bar{\psi}$, and so on.
There is only one 8-particle process that is not related by {\it (i)} to old cases. Working in the $x$ going to infinity limit, it is possible, however, to factorize this diagram into terms always containing some vanishing  ${\cal M}_{2\to 4}$. One can go on recursevely to higher point amplitudes, which will all vanish without imposing new constraints on the $\lambda_n$'s. 

In conclusion, we see that the choice of couplings in (\ref{SUSYEvenconstraintSL}) guarantees that no particle are produced in any process involving either bosons or fermions. The ${\cal N}=1$ supersymmetric sinh-Gordon potential (and its analytic continuation to sine-Gordon) is then the only supersymmetric model of a single scalar superfield with two supercharges and a $\mathbb{Z}_2$-symmetry that exists in two dimensions and does not produce particles. It would be interesting to extend this analysis to theories with more than one superfield and also to ${\cal N}=(2,2)$ multiplets.

We relax now the assumption of $\mathbb{Z}_2$-invariance. We discover that particle production does necessarily take place and apparently no supersymmetric extension of the Bullough-Dodd model exists. First of all, making use of (\ref{biglambda}) in (\ref{LambdanBD}), one can compute the first few $\lambda_n$'s:
\begin{equation}
    \lambda_4 = 6\lambda_3^2 \,,\qquad  \lambda_5 = 15\lambda_3^3 \,. 
    \label{BDchoice}
\end{equation}
In the fermionic sector, the simplest process is two fermions going into three bosons: $\bar{\psi}\psi \rightarrow \phi^3$. 
\begin{figure}
\begin{center}
    \includegraphics[scale=0.4]{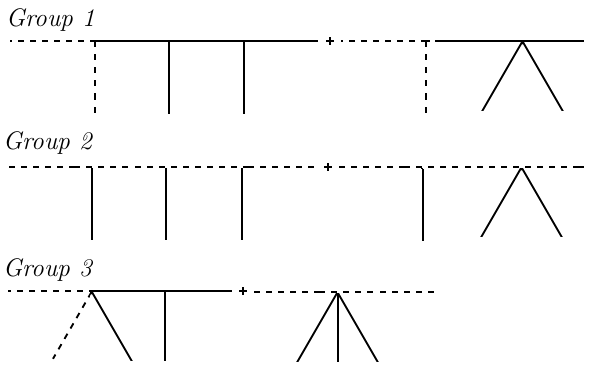}
    \end{center}
    \caption{\small Diagrams contributing to the process $\bar{\psi}\psi \rightarrow \phi\phi\phi$.}
    \label{SUSYM2f3b}
\end{figure}
There are six diagrams, contributing to the amplitude of this process, as shown in Fig.~\ref{SUSYM2f3b}. In the limit of interest for the momenta, all diagrams can be easily evaluated
\bea
 &&   \textrm{Group 1}\quad  \rightarrow\quad  \bar{v}_2\left( \lambda_3(-1)\Lambda_3(-1)\Lambda_3) + \lambda_3(-1)\Lambda_4 \right)u_1 = \bar{v_2}\left( -18\lambda_3^3 \right) u_1\,,\cr
  &&  \textrm{Group 2} \quad \rightarrow \quad 2\bar{v}_2 \left(3\lambda_3\lambda_4\slashed{q} \right)u_1 = \bar{v_2}\left( 36\lambda_3^3\slashed{q} \right) u_1 \,,\cr
  &&  \textrm{Group 3}\quad \rightarrow \quad \bar{v_2}\left( \lambda_4(-1)\Lambda_3 + \lambda_5 \right) u_1 = \bar{v_2}\left( \lambda_5 - 18\lambda_3^3 \right) u_1\,.
\eea
The amplitude of this process is
\begin{equation}
    {\cal M}_{\textrm{2f}\to \textrm{3b}} \rightarrow \bar{v_2}\left( 36\lambda_3^3\slashed{q} + \lambda_5 - 36\lambda_3^3 \right) u_1 = 6\lambda_3^3\bar{v}_2 \left( 6\slashed{q} + \beta \right)u_1\,,
\end{equation}
where
\begin{equation}
    \beta = \frac{1}{6\lambda_3^3}(\lambda_5 - 36\lambda_3^3)\,.
\end{equation}
This amplitude will be zero when $\beta = -3$, or
\begin{equation}
    \lambda_5 = 18\lambda_3^3\,,
\end{equation}
which is a different choice than the one in the bosonic case in (\ref{BDchoice}). Therefore, the bosonic and fermionic processes cannot be set to zero simultaneously with the same choice of couplings.

\subsubsection*{Acknowledgements}

We are happy to thank Joe Minahan and Martin Ro\v cek for comments. We are especially thankful to Pedro Vieira for the nice set of lectures he gave at the {\it XIX Swieca School on Particles and Fields}, which inspired this note. CB is grateful to Uppsala University for hospitality during this project. DT acknowledges partial financial support from CNPq and from the FAPESP grant 2014/18634-9. He also thanks ICTP-SAIFR for their support through the FAPESP grant 2016/01343-7.

\end{document}